\documentclass[twocolumn,showpacs,pra,aps,floatfix,superscriptaddress]{revtex4-1}
\usepackage[latin1]{inputenc}
\usepackage{bm}
\usepackage{multirow,amssymb,amsbsy,amsmath}
\usepackage{graphicx}
\makeatletter
\usepackage{pifont}
\makeatother
\newcommand{\tmtextit}[1]{{\itshape{#1}}}
\newcommand{\tmop}[1]{\ensuremath{\operatorname{#1}}}
\usepackage{color}

\begin{document}

\title{Initial correlations in open system's dynamics: The Jaynes-Cummings model}

\author{Andrea \surname{Smirne}}

\email{andrea.smirne@unimi.it}

\affiliation{Dipartimento di Fisica, Universit{\`a} degli Studi di
Milano, Via Celoria 16, I-20133 Milano, Italy}

\affiliation{INFN, Sezione di Milano, Via Celoria 16, I-20133
Milano, Italy}

\author{Heinz-Peter Breuer}

\email{breuer@physik.uni-freiburg.de}

\affiliation{Physikalisches Institut, Universit\"at Freiburg,
Hermann-Herder-Strasse 3, D-79104 Freiburg, Germany}

\author{Jyrki Piilo}

\email{jyrki.piilo@utu.fi}

\affiliation{Turku Centre for Quantum Physics, Department of
Physics and Astronomy, University of Turku, FI-20014 Turun
yliopisto, Finland}

\author{Bassano \surname{Vacchini}}

\email{bassano.vacchini@mi.infn.it}

\affiliation{Dipartimento di Fisica, Universit{\`a} degli Studi di
Milano, Via Celoria 16, I-20133 Milano, Italy}

\affiliation{INFN, Sezione di Milano, Via Celoria 16, I-20133
Milano, Italy}

\date{\today}

\begin{abstract}
Employing the trace distance as a measure for the
distinguishability of quantum states, we study the influence of
initial correlations on the dynamics of open systems. We
concentrate on the Jaynes-Cummings model for which the knowledge
of the exact joint dynamics of system and reservoir allows the
treatment of initial states with arbitrary correlations. As a
measure for the correlations in the initial state we consider the
trace distance between the system-environment state and the
product of its marginal states. In particular, we
examine the correlations contained in the thermal equilibrium
state for the total system, analyze their dependence on the
temperature and on the coupling strength, and demonstrate their
connection to the entanglement properties of the eigenstates of
the Hamiltonian. A detailed study of the time dependence of the
distinguishability of the open system states evolving from the
thermal equilibrium state and its corresponding uncorrelated
product state shows that the open system dynamically uncovers
typical features of the initial correlations.
\end{abstract}


\maketitle

\section{Introduction}
Quantum systems are typically subjected to the interaction with an
environment which influences their dynamics in a non negligible
way.  A realistic description, taking this external influence into
account, is crucial for the theoretical description of open quantum
systems, which play an important role in many areas of physics
\cite{Breuer2007}. The available theoretical tools allow a full
characterization for a Markovian dynamics, which can be described
by means of completely positive quantum dynamical semigroups
{\cite{Gorini1976a,Lindblad1976a}}. However, the assumptions
justifying the Markovian description of the open system dynamics
are often too restrictive, and a more general analysis is required.
A wealth of different approaches to deal with non-Markovian
dynamics have been introduced
\cite{Daffer2004a,Budini2004a,Shabani2005a,Maniscalco2006a,Vacchini2008a,Ferraro2008a,Kossakowski2008a,Kossakowski2009a,Breuer2008a,Breuer2009a,Piilo2008a,Piilo2009a,Breuer2009d,Breuer2009c,Laine2010a,Vacchini2010b,Smirne2010a,Chruscinski2010a,Chruscinski2010b,Mazzola2010a},
but they typically rely on the hypothesis that at the initial time
the open system and the environment are statistically independent.
This assumption is well justified in the case of weak interaction,
but in general one cannot neglect the initial correlations between
the open system and the environment
\cite{Uchiyama2010a,Pechukas1994a,Alicki1995a}.

Considering an open system $S$ which is coupled to an environment
$E$ and assuming that the composite system evolves according to a
unitary time evolution operator $U(t)$ from a total initial state
$\rho_{SE}(0)$, we can write the state of $S$ at time $t$
as follows,
\begin{equation}
 \rho_S(t) = \tmop{Tr}_E\left[U(t)\rho_{SE}(0)U^{\dagger}(t)\right].
 \label{eq:redmap}
\end{equation}
This equation defines a linear, completely positive and trace
preserving map $\Lambda_t$ from the state space of the total
system $S+E$ to the state space of the open system $S$:
\begin{equation}
 \rho_{SE}(0) \mapsto \rho_S(t) = \Lambda_t \rho_{SE}(0).
 \label{LAMBDA}
\end{equation}
If the open system $S$ and its environment $E$ are initially in an
uncorrelated tensor product state
\begin{equation} \label{INIT-PROD}
 \rho_{SE}(0) = \rho_S(0)\otimes\rho_E
\end{equation}
with a fixed environmental state $\rho_E$, Eq.~\eqref{eq:redmap}
also defines a linear map $\Phi_t$ from the state space of $S$
into itself,
\begin{equation} \label{CPT-MAP}
 \rho_S(0) \mapsto \rho_S(t) = \Phi_t \rho_S(0)
 = \tmop{Tr}_E\left[U(t) \rho_S(0)\otimes\rho_E U^{\dagger}(t)\right].
\end{equation}
It can be shown that this quantum dynamical map $\Phi_t$ is again
completely positive and trace preserving. Under the additional
assumption that the family of dynamical maps $\{\Phi_t, t\geq 0\}$
constitutes a semigroup, one derives the general mathematical
structure of its generator, which leads to the widely-used quantum
Markovian master equations for the open system state $\rho_S(t)$
in Lindblad form.

The above construction of the quantum dynamical map $\Phi_t$
presupposes that one restricts the class of initial conditions to
states of the form of Eq.~\eqref{INIT-PROD}, where $\rho_E$ is a
fixed, given environmental state. Thus, a large class of initial
conditions is excluded when considering dynamical maps acting on
the reduced state space, in particular those initial conditions
that describe correlations and entanglement between system and
environment. On the other hand, it is a well known fact that
correlations in the initial state can have strong influences on
the open system dynamics, both in thermal equilibrium and in
non-equilibrium systems. The question is thus, how do initial
correlations affect the reduced system dynamics, and what are
appropriate observable measures that quantify such effects? Here,
we discuss these questions in detail with the help of the example
of the Jaynes-Cummings model, the model of a two-state system
coupled to a bosonic field mode, employing the analytical
representation of the reduced dynamics of this model for arbitrary
initial states \cite{Smirne2010b}.

It should be mentioned that under certain additional assumptions
Eq.~\eqref{eq:redmap} can indeed be used to construct maps on the
reduced state space to represent the dynamics in the case of
initial correlations {\cite{vanWonderen2000a,Stelmachovic2001a,Jordan2004a,Rodriguez2008a,Carteret2008a,Shabani2009a}}. However, this construction demands that the
initial correlations between the system and its environment are
fixed. It turns out that the dynamical maps arising in this way
may be not completely positive, and not even positive. This
requires the determination of a certain compatibility domain in
the physical state space, which is a very complicated mathematical
task. In the present paper we shall follow an entirely different
strategy, to analyze the role of initial correlations. Namely, in
order to quantify the effect of initial system-environment
correlations in the subsequent time evolution of the open system,
we will investigate the trace distance
$D(\rho_S^1(t),\rho_S^2(t))$ between a pair of states
$\rho^1_S(t)$ and $\rho^2_S(t)$ of $S$, which evolve from a given
pair of initial states $\rho^1_{SE}(0)$ and
$\rho^2_{SE}(0)$ of the total system. This approach has
also been used in \cite{Breuer2009c,Laine2010a} to construct a
measure for the non-Markovianity of quantum processes, and in
\cite{Laine2010b} to develop a witness which allows the detection
of initial correlations through only local measurements on the
open system. An application to a specific system has been recently
considered in \cite{Dajka2010a}.

In the present paper we will address, in particular, the situation in
which $\rho_{SE}^1(0)$ represents a thermal equilibrium (Gibbs) state
corresponding to the full Hamiltonian of the model.  Since this
correlated state is invariant under the time evolution, its reduced
states $\rho_S^1(0)={\rm{Tr}}_E\rho_{SE}^1(0)$ and
$\rho_E^1(0)={\rm{Tr}}_S\rho_{SE}^1(0)$ remain of course time
independent. However, the initial state
$\rho_{SE}^2(0)=\rho_S^1(0)\otimes\rho_E^1(0)$ given by the product of
the marginals does evolve in time, and we will investigate the
dynamics of the trace distance $D(\rho_S^1(t),\rho_S^2(t))$ between
the open system states corresponding to the initial states
$\rho_{SE}^1(0)$ and $\rho_{SE}^2(0)$. For this case the trace
distance is bounded from above by the trace distance
$D(\rho^1_{SE}(0),\rho^1_S(0)\otimes\rho^1_E(0))$ which provides a
measure for the amount of correlations in the initial Gibbs state
\cite{Laine2010b}. Analyzing in detail the dependence on the
temperature and on the system-environment coupling strength, we
demonstrate that at small temperatures characteristic properties of
these correlations are related to the eigenvalue spectrum and, in
particular, to the quantum correlations and the entanglement structure
of the eigenstates of $H$. We will discuss further the signatures of
these properties in the subsequent dynamics of the open system states.
It will be shown that, in fact, the open system dynamically uncovers
typical features of the correlations in the initial states.

The paper is organized as follows. In
Sec.~\ref{sec:trace-dist-init} we introduce the trace distance and
show its relevance as a measure of the distinguishability of two
quantum states and of the correlations contained in a given
bipartite state. We further consider the exact reduced dynamics of
the Jaynes-Cummings model, and study as an example the time
behavior of the distinguishability of distinct initial states. In
Sec.~\ref{sec:gibbs} we provide a detailed study of the
correlations contained in the Gibbs state associated to the
Jaynes-Cummings Hamiltonian, as measured by the trace distance
between the state and the tensor product of its marginals. We then
investigate the time dependence of the distinguishability of the
corresponding time evolved states.

\section{Trace distance and initial system-environment correlations}
\label{sec:trace-dist-init}

\subsection{General theory}

\subsubsection{Properties and physical interpretation of the trace distance}

The trace distance of two trace class operators $A$ and $B$ is
defined as $\frac{1}{2}$ times the trace norm of $A-B$,
\begin{equation}
  D (A,B) = \frac{1}{2} ||A-B||_{1},
\end{equation}
where the trace norm of an operator $X$ is defined by
\begin{equation}
 ||X||_{1} = {\rm{Tr}} |X| = {\rm{Tr}} \sqrt{X^{\dagger}X}.
\end{equation}
If $X$ is  trace class and self-adjoint with eigenvalues $x_i$, this formula
reduces to the sum of the absolute eigenvalues (counting
multiplicity),
\begin{equation}
 ||X||_{1} = \sum_i |x_i|.
\end{equation}
The trace distance of two quantum states, represented by positive
operators $\rho^1$ and $\rho^2$ with unit trace, is thus given by
\begin{equation}
 D(\rho^1,\rho^2) = \frac{1}{2}||\rho^1-\rho^2||_{1}
 = \frac{1}{2} \tmop{Tr}|\rho^1-\rho^2|.
 \label{eq:trdist}
\end{equation}
The trace distance is a metric on the space of physical states
with several nice properties which make it a useful measure for
the distance between two quantum states. We list some of them:

\begin{description}

\item[1.] The trace distance for any pair of states satisfies the
inequality
\begin{equation}
 0 \leq D(\rho^1,\rho^2) \leq 1,
\end{equation}
where $D(\rho^1,\rho^2)=0$ if and only if $\rho^1=\rho^2$, and
$D(\rho^1,\rho^2)=1$ if and only if $\rho^1$ and $\rho^2$ have
orthogonal supports.

\item[2.] Being a metric, the trace distance satisfies the
triangular inequality,
\begin{equation} \label{TRIANGLE}
 D(\rho^1,\rho^2) \leq D(\rho^1,\rho^3) + D(\rho^3,\rho^2).
\end{equation}

\item[3.] All trace preserving positive maps $\Lambda$ are
contractions of the trace distance \cite{Ruskai1994a},
\begin{equation} \label{CONTRACTION}
 D(\Lambda\rho^1,\Lambda\rho^2) \leq D(\rho^1,\rho^2),
\end{equation}
where the equality sign holds if $\Lambda$ is a unitary
transformation.

\item[4.] The trace distance is subadditive with respect to the
tensor product,
\begin{equation} \label{SUB-1}
 D(\rho^1\otimes\sigma^1,\rho^2\otimes\sigma^2)
 \leq D(\rho^1,\rho^2) + D(\sigma^1,\sigma^2).
\end{equation}
In particular, one has
\begin{equation} \label{SUB-2}
 D(\rho^1\otimes\sigma,\rho^2\otimes\sigma) = D(\rho^1,\rho^2).
\end{equation}

\item[5.] The trace distance can be represented as a maximum taken
over all projection operators $\Pi$,
\begin{equation} \label{MAXI}
 D(\rho^1,\rho^2) = \max_{\Pi}
 {\rm{Tr}}\left\{\Pi\left(\rho^1-\rho^2\right)\right\}.
\end{equation}

\end{description}

The physical interpretation \cite{Gilchrist2005a} of the trace
distance is based on the relation \eqref{MAXI}. Suppose Alice
prepares a system in one of two quantum state $\rho^1$ and
$\rho^2$ with probability of $1/2$ each. She gives the system to
Bob, who performs a measurement in order to distinguish the two
states. Employing Eq.~\eqref{MAXI} one can show that the maximal
success probability for Bob to identify correctly the state is
given by $\left[1 + D(\rho^1,\rho^2)\right]/2$. This means that
the trace distance represents the maximal bias in favor of the
correct state identification which Bob can achieve through an
optimal strategy. Hence, the trace distance $D(\rho_1,\rho_2)$ can
be interpreted as a measure for the distinguishability of the
states $\rho^1$ and $\rho^2$.

\subsubsection{Dynamics of the trace distance}\label{Sec:D_Dyn}

We consider any two total system initial states $\rho^1_{SE}(0)$
and $\rho^2_{SE}(0)$, and the corresponding open system states
$\rho^1_S(t)$ and $\rho^2_S(t)$ at time $t$. According to
Eqs.~\eqref{eq:redmap} and \eqref{LAMBDA} the latter are given by
$\rho^1_S(t)=\Lambda_t\rho_{SE}^1(0)$ and
$\rho^2_S(t)=\Lambda_t\rho_{SE}^2(0)$. Since $\Lambda_t$ is
completely positive and trace preserving, we obtain from
Eq.~\eqref{CONTRACTION} a bound for the trace distance between the
reduced system states,
\begin{equation} \label{LAINE-0}
 D(\rho^1_S(t),\rho^2_S(t)) \leq
 D(\rho^1_{SE}(0),\rho^2_{SE}(0)).
\end{equation}
If the initial states are uncorrelated with the same environmental
state $\rho_E$, that is
$\rho^1_{SE}(0)=\rho_S^1(0)\otimes\rho_E(0)$ and
$\rho^2_{SE}(0)=\rho_S^2(0)\otimes\rho_E(0)$, this
inequality reduces with the help of \eqref{SUB-2} to the
contraction property for the dynamical map \eqref{CPT-MAP},
\begin{equation}
 D(\rho^1_S(t),\rho^2_S(t)) \leq
 D(\rho^1_S(0),\rho^2_S(0)).
\end{equation}
This means that for initially uncorrelated total system states and
identical environmental states the trace distance between the
reduced system states at time $t$ can never be larger than its
initial value.

The inequality \eqref{LAINE-0} may be written as
\begin{eqnarray}
 &&D(\rho^1_S(t),\rho^2_S(t)) - D(\rho^1_S(0),\rho^2_S(0))\nonumber\\
 &&\leq  D(\rho^1_{SE}(0),\rho^2_{SE}(0))
 - D(\rho^1_S(0),\rho^2_S(0))\nonumber\\&& \equiv I(\rho^1_{SE}(0),\rho^2_{SE}(0)).
 \label{eq:laine}
\end{eqnarray}
According to this inequality the change of the trace distance of
the open system states is bounded from above by the quantity
$I(\rho^1_{SE}(0),\rho^2_{SE}(0))\geq 0$. This quantity represents
the distinguishability of the total system initial states minus
the distinguishability of the corresponding reduced system initial
states. Thus, $I(\rho^1_{SE},\rho^2_{SE})$ can be interpreted as
the relative information of the total initial states which is
initially outside the open system, i.e., which is inaccessible for
local measurement performed on the open system \cite{Laine2010b}.

For $I(\rho^1_{SE}(0),\rho^2_{SE}(0))>0$ the trace distance of the
open system states can increase over its initial value. This
increase can be interpreted by saying that information which is
initially outside the open system flows back to the system and
becomes accessible through local measurements. Note that, as will
be illustrated by means of several examples below, the bound for
the dynamics of the trace distance given by Eq.~(\ref{eq:laine})
is tight, i.e., it can be reached for certain total initial
states. If the bound of inequality (\ref{eq:laine}) is actually
reached at some time $t$, the initial distinguishability of the
total system states is equal to the distinguishability of the open
system states at time $t$. This means that the relative
information on the total initial states has been dynamically
transferred completely to the open system \cite{Laine2010b}.

Using the sub-additivity of the trace distance \eqref{SUB-1} and
the triangular inequality \eqref{TRIANGLE} one deduces from
(\ref{eq:laine}) the following inequality \cite{Laine2010b},
\begin{eqnarray} \label{general-inequality}
&& D(\rho_S^1(t),\rho_S^2(t)) - D(\rho_S^1(0),\rho_S^2(0))\nonumber\\
 &&\leq D(\rho_{SE}^1(0),\rho_S^1(0)\otimes\rho_E^1(0))\nonumber\\
 &&+ D(\rho_{SE}^2(0),\rho_S^2(0)\otimes\rho_E^2(0)) + D(\rho_E^1(0),\rho_E^2(0)).
\end{eqnarray}
For any state $\rho_{SE}$ the quantity
$D(\rho_{SE},\rho_S\otimes\rho_E)$ describes how well $\rho_{SE}$
can be distinguished from the fully uncorrelated product state
$\rho_S\otimes\rho_E$ of its marginal states $\rho_S$ and
$\rho_E$. Thus, $D(\rho_{SE},\rho_S\otimes\rho_E)$ can be
interpreted as a measure for the total amount of correlations in
the state $\rho_{SE}$. Therefore, the inequality
\eqref{general-inequality} shows that an increase of the trace
distance of the open system states over its initial value implies
that there must be correlations in the initial states
$\rho_{SE}^1(0)$ or $\rho_{SE}^2(0)$, or that the environmental
states are different. An important special case, which will be
considered in detail in the present paper, occurs if
$\rho^2_{SE}(0)$ is given by the product state obtained from the
marginals of $\rho^1_{SE}(0)$, i.e.,
$\rho^2_{SE}(0)=\rho^1_S(0)\otimes\rho^1_E(0)$. The inequality
(\ref{eq:laine}) then reduces to the simple form
\begin{equation}
 D(\rho^1_S(t),\rho^2_S(t)) \leq D(\rho^1_{SE}(0),\rho^1_S(0)\otimes\rho^1_E(0)),
 \label{eq:llaine}
\end{equation}
according to which the increase of the trace distance is bounded
by the amount of correlations in the total initial state \cite{Laine2010b}.

\subsection{Example: The Jaynes-Cummings model}
\label{sec:ex}

\subsubsection{The physical model}
We consider a two-state system coupled to a single mode of the
radiation field with total Hamiltonian
\begin{eqnarray}
  H &=& H_S + H_E + H_I\nonumber\\ &=& \omega_0 \sigma_+ \sigma_- + \omega b^{\dag} b + g
  \left( \sigma_+ \otimes b + \sigma_- \otimes b^{\dag} \right),  \label{eq:h}
\end{eqnarray}
where $\sigma_+ = |1 \rangle \langle 0|$ and $\sigma_- = |0
\rangle \langle 1|$ are the raising and lowering operators of the
two-state system, $b^{\dag}$ and $b$ are the creation and
annihilation operators of the field mode, and the coupling term is
in the Jaynes-Cummings form. This model describes, e.g., the
interaction between a two-level atom and a mode of the radiation
field in the electric dipole and rotating wave approximation. In
the interaction picture the Hamiltonian takes the form
\begin{equation}
  H_I (t) = g \left( \sigma_+ \otimes b e^{i \Delta t} + \sigma_- \otimes
  b^{\dag} e^{- i \Delta t} \right),  \label{eq:hint}
\end{equation}
where $\Delta=\omega_0-\omega$ denotes the detuning between the
system's transition frequency $\omega_0$ and the frequency
$\omega$ of the field mode. The exact time-evolution operator for
the total system in the interaction picture can then be written as
(see, e.g., Ref.~{\cite{Puri2001}}):
\begin{equation}
  U(t) = \left(\begin{array}{cc}
    c_{} ( \hat{n} + 1, t) & d_{} ( \hat{n} + 1, t) b\\
    - b^{\dag} d^{\dag} ( \hat{n} + 1, t) & c^{\dag} ( \hat{n}, t)
  \end{array}\right),  \label{eq:U}
\end{equation}
where we have introduced the following functions of the number
operator $\hat{n} = b^{\dag}b$,
\begin{eqnarray}
  c ( \hat{n}, t) & = & e^{i \Delta t / 2}  \left[ \cos \left( \Omega (
  \hat{n})_{} \frac{t}{2} \right) - i \frac{\Delta}{\Omega ( \hat{n})} \sin
  \left( \Omega ( \hat{n}) \frac{t}{2} \right) \right] ,\nonumber\\
  d ( \hat{n}, t) & = & - ie^{i \Delta t / 2} \frac{2 g}{\Omega ( \hat{n})}
  \sin \left( \Omega ( \hat{n}) \frac{t}{2} \right),  \label{eq:cd}
\end{eqnarray}
with
\begin{equation}
  \Omega (\hat{n}) = \sqrt{\Delta^2 + 4 g^2 \hat{n}}.  \label{eq:omn}
\end{equation}

With the help of the unitary time-evolution operator given by
Eq.~(\ref{eq:U}) we can easily determine the exact expression for
the reduced density matrix of the two-level system at time $t$,
\begin{equation}
  \rho_S (t) = \left(\begin{array}{cc}
    \rho_{11} (t) & \rho_{10} (t)\\
    \rho^{\ast}_{10} (t) & \rho_{00} (t)
  \end{array}\right),  \label{eq:rhost}
\end{equation}
corresponding to an arbitrary initial state $\rho_{SE}(0)$
of the total system. First, we expand $\rho_{SE}(0)$ with
respect to the basis vectors $|\alpha \rangle \otimes |n \rangle
\equiv | \alpha, n \rangle$, where $\alpha = 1,0$ labels the
states of the two-state system, and $n=0,1,2,\ldots$ the number
states of the field mode,
\begin{equation}
  \rho_{SE} (0) = \sum_{\alpha, \beta, m, n} \rho^{mn}_{\alpha
  \beta} (0) | \alpha, m \rangle \langle \beta, n|.  \label{eq:rseo}
\end{equation}
Substituting this expression into Eq.~(\ref{eq:redmap}) with
$U(t)$ given by Eq.~(\ref{eq:U}), one obtains
\begin{eqnarray}
&&  \rho_{11}(t) =  \sum_n \left[ \rho^{nn}_{11} (0) |c_{n + 1} (t)
  |^2 + 2 \sqrt{n + 1} \right.\nonumber\\
&& \left. \times \tmop{Re} \left\{ \rho^{n,n+1}_{10}(0)
  d^{\ast}_{n + 1} (t) c_{n + 1} (t) \right\} + n \rho^{nn}_{00} (0) |d_n
  (t) |^2 \right] \nonumber\\
 && \rho_{10}(t)  =  \sum_n \left[ - \sqrt{n + 1} \rho^{n+1,n}_{11}
  (0) c_{n + 2} (t) d_{n + 1} (t) \right.\nonumber\\
 && \left.- \sqrt{n + 2} \sqrt{n + 1} \rho^{n+2,
  n}_{01} (0) d_{n + 2} (t) d_{n + 1} (t) \right. \nonumber\\
  &  & \left. + \rho^{nn}_{10} (0) c_{n + 1} (t) c_n (t) + \sqrt{n + 1}
  \rho^{n+1,n}_{00} (0) d_{n + 1} (t) c_n (t) \right],\nonumber\\&&  \label{eq:eqvt}
\end{eqnarray}
where $c_n(t)$ and $d_n(t)$ denote the eigenvalues of
$c(\hat{n},t)$ and $d(\hat{n},t)$ corresponding to the eigenstate
$|n \rangle$, respectively.

We note that Eq.~\eqref{eq:eqvt} does in general not lead to a
dynamical map for the state changes of the reduced two-state
system since it is not possible to write the right-hand side of
this equation as a function of the matrix elements of the reduced
initial state $\rho_S (0)$ which are given by
\begin{equation}
 \rho_{\alpha\beta}(0) = \sum_n \rho^{nn}_{\alpha\beta}(0).
 \label{eq:rhoab0}
\end{equation}
However, if the total initial state is of tensor product form,
$\rho_{SE}(0)=\rho_S(0)\otimes\rho_E(0)$ and, therefore,
\begin{equation}
 \rho^{nm}_{\alpha\beta} (0) = \rho_{\alpha\beta} (0)\rho^{nm}(0),
 \label{eq:rhoab02}
\end{equation}
it is indeed possible to construct the dynamical map; if moreover
$\left[\rho_E (0),\hat{n}\right]=0$, one finds the map already
derived in Ref.~\cite{Smirne2010b}.

\subsubsection{Dynamics of the trace distance for pure or product total initial states}

We illustrate the dynamics of the trace distance and the
inequality (\ref{eq:laine}) by means of two simple examples,
considering the situation in which the total initial state is a
product state or a pure state. The case of a mixed, correlated
initial state will be considered in detail in
Sec.~\ref{sec:gibbs}.

\begin{figure}[h]
\begin{center}
\includegraphics[scale=0.53]{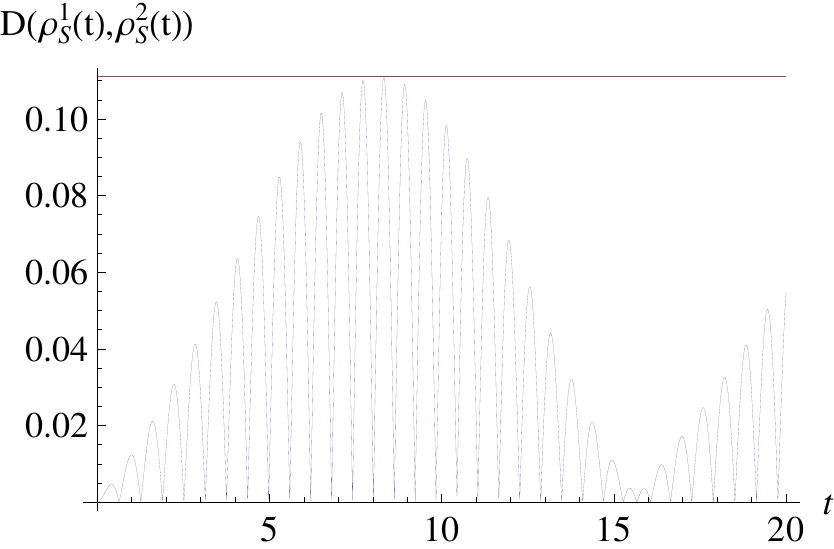}\includegraphics[scale=0.53]{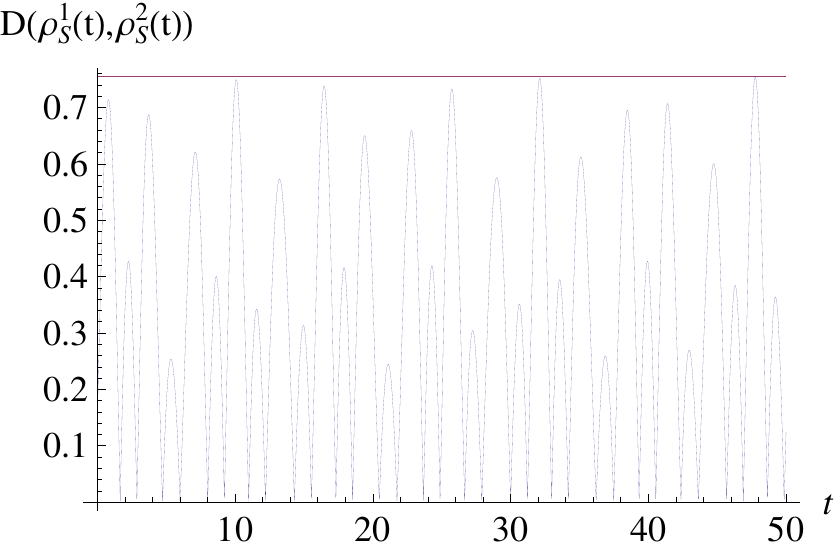}
\caption{\label{fig:1}(Color online) Plot of the trace
distance $D(\rho^1_S(t),\rho^2_S(t))$ as a function of time, where
$\rho_S^1(t)$ and $\rho_S^2(t)$ have been determined from
Eq.~(\ref{eq:eqvt}). In both figures the horizontal line marks the
upper bound of Eq.~(\ref{eq:laine}), and $\Delta = 0.1, g = 1$
(arbitrary units). (a) Dynamics for two product total initial
states which differ only by the environmental states and are given
by Eq.~(\ref{eq:1}) and (\ref{eq:2}) with $|\alpha_1|^2=7/9$,
$|\alpha_2|^2=8/9$ and $n = 7$. (b) The two reduced states
$\rho^1_S (t)$ and $\rho^2_S (t)$ are obtained from the total
initial states given by Eqs.~(\ref{eq:3}) and (\ref{eq:4}) which
have the same environmental marginal state, but different reduced
system states and correlations. Parameters: $\alpha=i\sqrt{3/7}$,
$\beta=\sqrt{4/7}$ and $n=1$.}
\end{center}
\end{figure}

The quantity on the right-hand side of Eq.~(\ref{eq:laine}),
representing the information which is initially outside the
reduced system, can be larger than zero basically for two reasons:
First, because one has different environmental initial states
$\rho^1_E (0)$ and $\rho^2_E (0)$ and, second, because of the
presence of correlations in the initial states
$\rho^1_{SE}(0)$ or $\rho^2_{SE}(0)$ (see inequality
\eqref{general-inequality}). To illustrate the first case we study
the trace distance between the two reduced states $\rho^1_S (t)$
and $\rho^2_S (t)$ evolving from two product initial states with
the same reduced system state, namely from
$\rho_{SE}^1(0)=\rho_S(0)\otimes\rho^1_E(0)$ and
$\rho_{SE}^2(0)=\rho_S (0)\otimes \rho^2_E (0)$, where
\begin{equation}
   \label{eq:1}
 \rho^{}_S (0)=| \alpha_1 |^2
 |0 \rangle \langle 0| + | \beta_1 |^2 |1 \rangle \langle 1|
\end{equation}
and the two environmental states are taken to be
\begin{equation}
   \label{eq:2}
   \rho^i_E (0)=| \alpha_i |^2 |n \rangle \langle n| +  | \beta_i |^2 |n - 1 \rangle
  \langle n - 1|, \qquad i=1,2,
\end{equation}
with the normalization condition $|\alpha_i|^2+|\beta_i|^2=1$.
Numerical simulation results for this case are shown in
Fig.~(\ref{fig:1}.a). We see from the figure that the bound of
Eq.~(\ref{eq:laine}), which is given by
$\left| |\alpha_1|^2-|\alpha_2|^2\right|$, is indeed reached here.
For a study of the second case we consider an initially correlated
pure state of the form
\begin{equation}
   \label{eq:3}
  \rho^1_{SE} (0) = | \psi \rangle
\langle \psi | ,
\end{equation}
with $|\psi\rangle=\alpha|0,n\rangle + \beta|1,n-1\rangle$,
$|\alpha|^2+|\beta|^2=1$, together with an initial product state
of the form
\begin{equation}
   \label{eq:4}
   \rho^2_{SE}
(0) =\rho^{2}_S (0) \otimes \rho^2_E (0)
\end{equation}
with $\rho^2_{S} (0) = | \beta |^2 |0 \rangle \langle 0| + |
\alpha_{} |^2 |1 \rangle \langle 1|$ and $\rho^2_{E} (0) =|
\alpha_{} |^2 |n \rangle \langle n| + | \beta |^2 |n - 1 \rangle
\langle n - 1|$. Note that $\rho^2_{SE}(0)$ is not equal to
the product of the marginals of $\rho^1_{SE}(0)$. As can be
seen from Fig.~(\ref{fig:1}.b) also for this case the bound of
Eq.~(\ref{eq:laine}), which is given by $\frac{1}{2}(1+|\alpha|^4
+|\beta|^4)$, is repeatedly reached in the course of time. As
expected, in both cases the trace distance of the states exceeds
its initial value, corresponding to the fact that the reduced
system dynamically retrieves the information initially not
accessible to it, related to the different initial environmental
states or to the initial system-environment correlations.
Note that the trace distance starts increasing already at the initial
time, indicating that the information is flowing to the reduced system
from the very beginning of the dynamics. Moreover, it keeps
oscillating also for large values of $t$, so that the
distinguishability growth between reduced states can be detected,
e.g. by quantum state tomography, also making observations after a
long interaction time.

In both situations considered and visualized in Fig.~(\ref{fig:1})
the maximum value of the trace distance as a function of time is
equal to the upper bound given by Eq.~(\ref{eq:laine}), indicating
that the information initially inaccessible to the reduced system
has been transferred completely to it during the subsequent
dynamics. This is of course not always the case and it is an
important problem to characterize explicitly those initial states
for which such a behavior indeed occurs. Let us consider the
special case given by Eq.~(\ref{eq:llaine}), in which the two
total initial states are a correlated state and the tensor product
of its marginals, taking $\rho^1_{SE}(0)$ to be a pure entangled
state, i.e., $\rho^1_{SE}(0)=|\psi\rangle\langle\psi|$ with $|\psi
\rangle=\alpha|0,n\rangle+\beta|1,m\rangle$. For this case
Eq.~(\ref{eq:eqvt}) leads to
\begin{eqnarray}
  &&D (\rho^1_S (t), \rho^2_S (t)) = \left| | \alpha \beta |^2 (|c_{m + 1} (t) |^2
  - |c_{n} (t) |^2+|c_{m} (t) |^2\right.\nonumber\\
 && \left. - |c_{n+1} (t) |^2) + 2 \delta_{m, n - 1} \sqrt{n} \tmop{Re} \left\{
  \alpha^{\ast} \beta d^{\ast}_n (t) c_n (t) \right\}\right|, \label{eq:drrpure}
\end{eqnarray}
while the right-hand side of Eq.~(\ref{eq:llaine}) becomes
\begin{equation}
  D \left( \rho^1_{SE} (0), \rho^1_S (0) \otimes \rho^1_E (0) \right)
  = | \alpha \beta |^2 + | \alpha \beta |.  \label{eq:limrrpure}
\end{equation}
Taking into account Eq.~(\ref{eq:cd}) and Eq.~\eqref{eq:omn},  for $n,m \gg \Delta^2/ 4g^2$
Eq.~(\ref{eq:drrpure}) explicitly reads
\begin{eqnarray}
  D (\rho^1_S (t), \rho^2_S (t))  &=&  \Big| | \alpha \beta |^2 \left[
  \cos^{2} \left( g \sqrt{m + 1} t \right) - \cos^{2} \left(g \sqrt{n}
  t \right)\right.\nonumber\\&&\left. +\cos^{2} \left( g \sqrt{m}t \right) - \cos^{2} \left(g \sqrt{n+1}
  t \right) \right] \nonumber \\
  && -  \delta_{m, n - 1} \tmop{Im} \left\{ \alpha^{\ast} \beta
  \right\} \sin \left( 2 g \sqrt{n} t \right) \Big|,   \label{eq:drrbign}
\end{eqnarray}
which is an almost periodic function \cite{Corduneanu1989} since
it represents a linear combination of sine and cosine functions
with incommensurable periods. The supremum of the attained values
\cite{Note1} is less than or equal to $2|\alpha\beta|^2$ if $m\neq n$ and
$m\neq n-1$, and equal to
$|\alpha\beta|^2+|\tmop{Im}\left\{\alpha^{\ast}\beta\right\}|$ if
$m=n-1$. Thus, the inequality in Eq.~(\ref{eq:llaine}) is tight
only for those initial states for which $m=n-1$ and $\tmop{Re}
\left\{\alpha^{\ast}\beta\right\}=0$ (indeed, we have
$|\alpha\beta|^2+|\alpha\beta|=2|\alpha\beta|^2$ if and only if
either $\alpha=0$ or $\beta=0$). The special role of the initial
states with $m=n-1$ can be traced back to the structure of the
full unitary evolution given by Eq.~(\ref{eq:U}) and to the
presence of the creation and annihilation operators in the
off-diagonal matrix elements.  Their action generates, in fact,
the last term in the modulus on the right-hand side of
Eq.~(\ref{eq:drrbign}), which for $m=n-1$ is necessary to reach
the bound.  If the relation $n,m \gg \Delta^2/ 4g^2$ is not
satisfied, the supremum lies in general strictly below the bound
even if the above mentioned conditions are fulfilled.  This is a
consequence of the fact that the periodic functions $|c_n(t)|^2$
are then strictly less than 1. 

\section{Gibbs initial state and dynamics of the trace distance}
\label{sec:gibbs}

We now extend our considerations to the evolution of the trace
distance between a mixed correlated initial state and the tensor
product of its marginals. Specifically, we will analyze the
inequality given in Eq.~(\ref{eq:llaine}) when the correlated
initial state $\rho_{SE}$ is the invariant Gibbs (thermal
equilibrium) state corresponding to the full Hamiltonian $H$ of
the model. For simplicity we will omit in the following the time
argument zero. We first analyze the total amount of correlations
in the initial state
$D\left(\rho_{SE},\rho_S\otimes\rho_E\right)$, i.e., the
upper bound for the trace distance according to
Eq.~(\ref{eq:llaine}). As we shall show below the main features of
this bound can be explained in terms of the correlations in the
ground state of the Hamiltonian $H$. We further study the behavior
of the actual dynamics of the trace distance, which will turn
out to reflect the characteristic features of the correlations in
the Gibbs state.

\subsection{Correlations in the Gibbs state}

We consider the total initial Gibbs state
\begin{equation}
 \rho_{SE} = \frac{1}{Z} e^{-\beta H},  \label{eq:gibbs}
\end{equation}
where $H$ is the total Hamiltonian of the system given by
Eq.~(\ref{eq:h}), $Z=\tmop{Tr}e^{-\beta H}$ denotes the partition
function and $\beta=1/k_bT$ with $k_b$ the Boltzmann constant and
$T$ the temperature. To calculate the marginal states $\rho_S =
\tmop{Tr}_E e^{- \beta H} / Z$ and $\rho_E = \tmop{Tr}_S e^{-
\beta H} / Z$ it is useful to obtain the matrix elements of
$\rho_{SE}$ with respect to the basis $\left\{ | \alpha, n \rangle
\right\}$ already introduced in Sec.~\ref{sec:ex}.
This can be done using the dressed states {\cite{Meystre1991a}},
i.e., the eigenvectors of the Hamiltonian $H$. These eigenvectors
can be written as
\begin{eqnarray}
  | \Phi^+_n \rangle & = & a_n |1, n - 1 \rangle + b_n |0, n \rangle,
  \hspace{2em}\nonumber\\
  | \Phi^-_n \rangle & = & - b_n |1, n - 1 \rangle + a_n |0, n \rangle,
  \hspace{1em}\nonumber\\
   | \Phi^-_0 \rangle&=&|0, 0 \rangle, \label{eq:dressed}
\end{eqnarray}
with $ n = 1,2,3,\ldots $ and
\begin{equation}
  a_n = \sqrt{\frac{\Omega_n + \Delta}{2 \Omega_n}}, \hspace{2em} b_n =
  \sqrt{\frac{\Omega_n - \Delta}{2 \Omega_n}},  \label{eq:sincos}
\end{equation}
where $\Omega_n=\sqrt{\Delta^2+4g^2n}$ (see
Eq.~\eqref{eq:omn}). The corresponding eigenvalues are given by
\begin{eqnarray}
 E^{\pm}_n &=& n \omega + \frac{\Delta}{2} \pm \frac{\Omega_n}{2}, \nonumber \\
 E^-_0 &=& 0.  \label{eq:eigenvalues}
\end{eqnarray}
Inverting Eqs.~(\ref{eq:dressed}) with the help of the relations
\begin{eqnarray}
  |0, n \rangle & = & b_n | \Phi^+_n \rangle + a_n | \Phi^-_n \rangle,
  \nonumber\\
  |1, n \rangle & = & a_{n + 1} | \Phi^+_{n + 1} \rangle - b_{n + 1} |
  \Phi^-_{n + 1} \rangle,  \label{eq:dressedinv}
\end{eqnarray}
one obtains the expressions
\begin{eqnarray}
  \rho^{nm}_{00} & = & \frac{1}{Z} \delta_{n, m} \left( e^{- \beta E^+_n}
  b_n^2 + e^{- \beta E^-_n} a_n^2 \right), \nonumber\\
  \rho^{nm}_{11} & = & \frac{1}{Z} \delta_{n, m} \left( e^{- \beta E^+_{n
  + 1}} a_{n + 1}^2 + e^{- \beta E^-_{n + 1}} b_{n + 1}^2 \right), \nonumber\\
  \rho^{nm}_{10} = \rho^{mn}_{01} & = & \frac{1}{Z} \delta_{n + 1, m}
  \left( e^{- \beta E^+_{n + 1}} - e^{- \beta E^-_{n + 1}} \right) a_{n + 1}
  b_{n + 1}, \nonumber\\&& \label{eq:gibbsmat}
\end{eqnarray}
which represent the matrix elements of the Gibbs state,
\begin{equation}
   \label{eq:100}
   \rho_{SE} = \sum_{\alpha,\beta,n,m} \rho^{nm}_{\alpha\beta}
   |\alpha,n\rangle\langle\beta,m|.
\end{equation}
Using this result together with Eq.~(\ref{eq:rhoab0}) we see that
the reduced system state is diagonal in the basis
$|\alpha,n\rangle$ and that the diagonal elements are given by
$\rho_{11}=1-\rho_{00}$ and
\begin{equation} \label{eq:10}
 \rho_{00} =  \frac{1}{Z} \sum_{n = 0}^{\infty} \left( e^{-\beta E^+_n}
  b_n^2 + e^{- \beta E^-_n} a_n^2 \right).
\end{equation}
The reduced state of the environment is also diagonal since
$\rho^{nm}=0$ for $n \neq m$, and the diagonal elements can be
expressed as
\begin{eqnarray}
  \rho^{nn}& = & \frac{1}{Z} \left( e^{- \beta E^+_n} b_n^2 + e^{- \beta
  E^-_n} a_n^2 \right.\nonumber\\&&\left.+ e^{- \beta E^+_{n + 1}} a_{n + 1}^2 + e^{- \beta E^-_{n + 1}}
  b_{n + 1}^2 \right).  \label{eq:red}
\end{eqnarray}
The product state constructed from the marginals is accordingly of
the form
\begin{eqnarray}
  \rho_S \otimes \rho_E & = & \sum_{\alpha,n} \rho_{\alpha\alpha} \rho^{nn}
  | \alpha, n \rangle \langle \alpha, n|.  \label{eq:prodct}
\end{eqnarray}
Finally, the normalization constant $Z$ can be written as
\begin{eqnarray}
  Z & = & \sum_n \left( e^{- \beta E^+_n} b_n^2 + e^{- \beta E^-_n} a_n^2 \right.\nonumber\\&&+
  \left.e^{- \beta E^+_{n + 1}} a_{n + 1}^2 + e^{- \beta E^-_{n + 1}} b_{n + 1}^2
  \right) .  \label{eq:Z}
\end{eqnarray}

Starting from the above relations we can analytically calculate
the total amount of correlations of the Gibbs state, i.e., the
quantity $D(\rho_{SE},\rho_S\otimes\rho_E)$. To this
end, we order the elements of the basis as $\left\{ |0, 0 \rangle,
|1, 0 \rangle, |0, 1 \rangle, |1, 1 \rangle, |0, 2 \rangle, |1, 2
\rangle, \ldots \right\}$. The difference
$X=\rho_{SE}-\rho_S\otimes\rho_E$ between the Gibbs state and its
corresponding product state can then be written in block diagonal
form,
\begin{equation}
  X =
  \left(\begin{array}{cccccccc}
    D^0_0 & 0 & 0 & 0 & 0 & 0 & \ldots & \ldots\\
    0 & D^0_1 & \rho^{01}_{10} & 0 & 0 & 0 & \ldots & \ldots\\
    0 & \rho^{10}_{01} & D^1_0 & 0 & 0 & 0 & \ldots & \ldots\\
    0 & 0 & 0 & D^1_1 & \rho^{12}_{10} & 0 & \ldots & \ldots\\
    0 & 0 & 0 & \rho^{21}_{01} & D^2_0 & 0 & \ldots & \ldots\\
    0 & 0 & 0 & 0 & 0 & \ddots & 0 & 0\\
    \vdots & \vdots & \vdots & \vdots & \vdots & D^n_1 & \rho^{n,n+1}_{10} & 0\\
    \vdots & \vdots & \vdots & \vdots & \vdots & \rho^{n+1,n}_{01} & D^{n
    + 1}_0 & 0\\
    \vdots & \vdots & \vdots & \vdots & \vdots & 0 & 0 & \ddots
  \end{array}\right),  \label{eq:megamatrix}
\end{equation}
where
\begin{eqnarray}
  D^n_{\alpha} & = & \rho^{n, n}_{\alpha, \alpha} - \rho_{\alpha, \alpha}
  \rho^{n, n} .  \label{eq:d}
\end{eqnarray}
It is easy to demonstrate that $D^n_1=-D^n_0$, implying that the
matrix of Eq.~(\ref{eq:megamatrix}) has zero trace, as it should
be. The eigenvalues of this matrix are simply given by the
eigenvalues of the $2\times 2$ block matrices plus the top left
element $D_0^0$. Hence, the total amount of correlations in the
Gibbs state is given by
\begin{eqnarray}
  &&D \left( \rho^{}_{SE}, \rho^{}_S \otimes \rho^{}_E \right) =
  \frac{1}{2} |D^0_0 | \nonumber\\&&+ \frac{1}{4} \sum_{n = 0}^{\infty} \left| D^n_1 + D^{n +
  1}_0+ \sqrt{\left( D^n_1 - D^{n + 1}_0 \right)^2 + 4 \left( \rho^{n, n +
  1}_{1, 0} \right)^2} \right|   \nonumber\\
  &  & + \frac{1}{4} \sum_{n = 0}^{\infty} \left| D^n_1+ D^{n + 1}_0-
  \sqrt{\left( D^n_1 - D^{n + 1}_0 \right)^2 + 4 \left( \rho^{n, n + 1}_{1, 0}
  \right)^2} \right| .\nonumber\\&&  \label{eq:tdgibbs}
\end{eqnarray}
This quantity depends on the model parameters $\omega$, $\Delta$
and $g$ which characterize the Hamiltonian described by
Eq.~(\ref{eq:h}), as well as on the temperature. In the following
we will focus in particular on the dependence of
$D(\rho_{SE},\rho_S\otimes\rho_E)$ on the coupling constant
$g$ and on the inverse temperature $\beta$ for fixed values of the
other two parameters (indeed from the expression of the Gibbs
state it immediately appears that the dependence on one of the
parameters can be reabsorbed into the others).

\subsection{Dependence on the ground state}\label{sec:conn-ground-level}

\begin{figure}[h]
\includegraphics[scale=0.47]{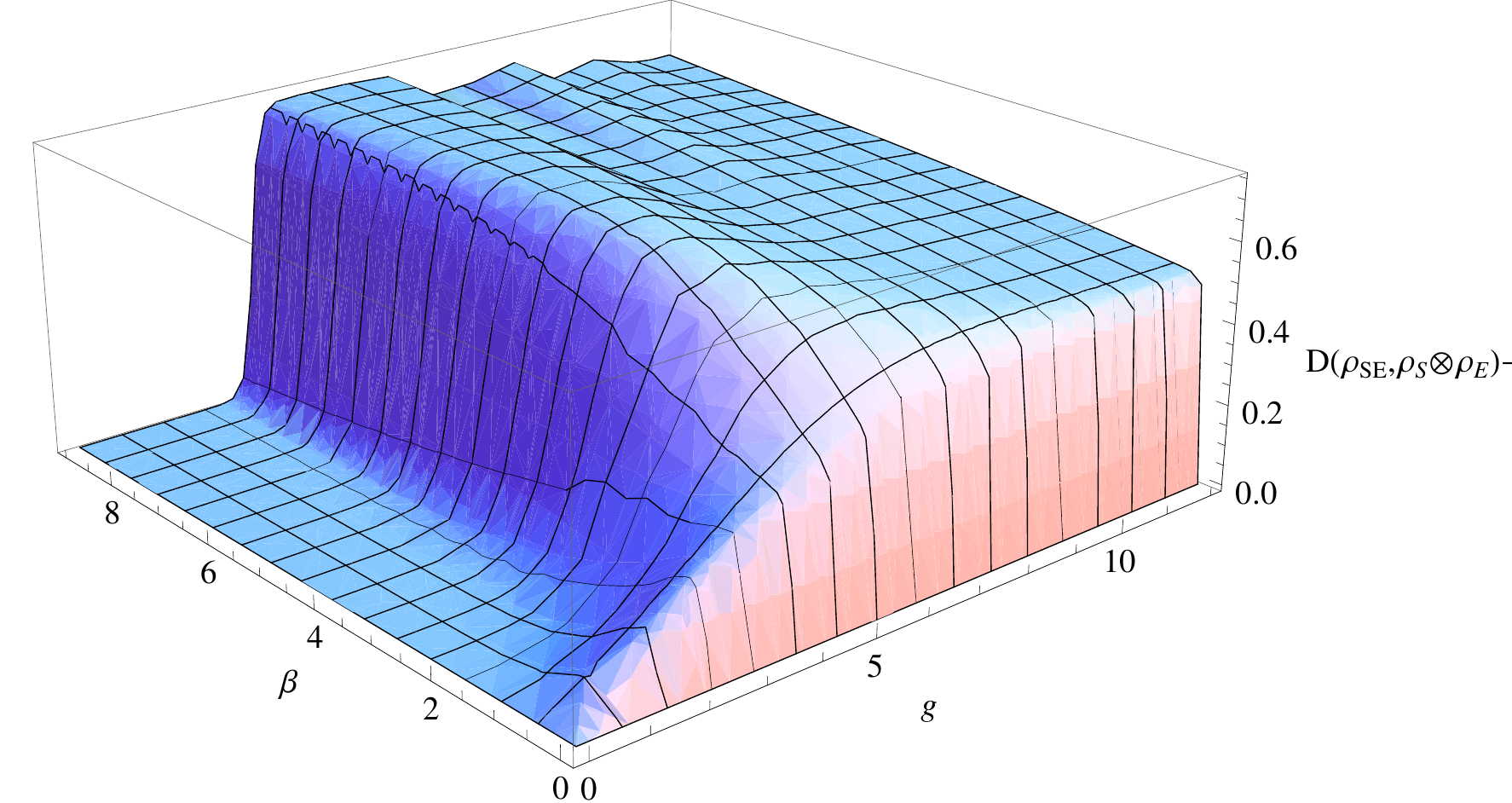}
\caption{\label{fig:2} (Color online) Plot of the
correlations of the Gibbs state \eqref{eq:gibbs} as a function of
the inverse temperature $\beta$ and of the coupling constant $g$
according to Eq.~(\ref{eq:tdgibbs}). Parameters: $\omega = 3$ and
$\Delta = 0.5$.}
\end{figure}

\begin{figure}[h]
\includegraphics[scale=0.65]{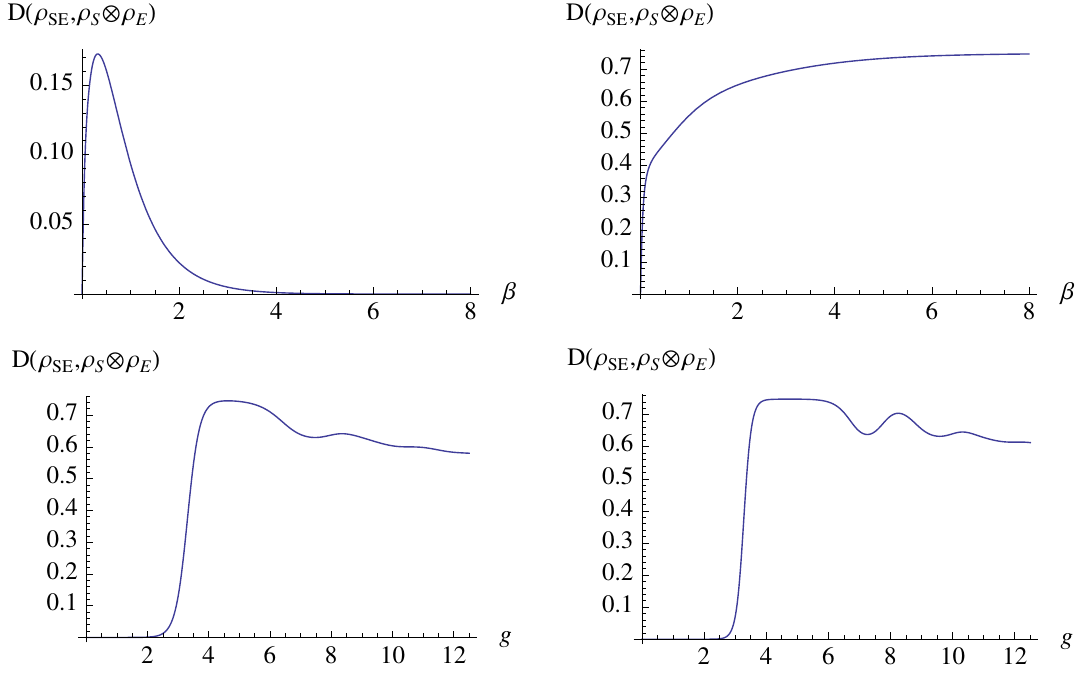}
\caption{\label{fig:3} (Color online) (a, b, c, d From top left to
bottom right) Sections of the plot in Fig.~(\ref{fig:2})
corresponding to $g = 1.7$, $g = 5.5$ , $\beta = 5$ and $\beta =
8$, respectively. The critical value of $g$ is given
by $\bar{g}_1=3.24$, see Eq.~(\ref{eq:criticalg}).}
\end{figure}

The behavior of the trace distance given by Eq.~(\ref{eq:tdgibbs}) as
a function of $\beta$ and $g$ is plotted in Fig.~(\ref{fig:2}). We
clearly see a non-monotonic behavior of the trace distance as a
function of both parameters.  Focusing on the dependence on $\beta$
for a fixed value of $g$, we observe that there is a sudden transition
between two different kinds of behavior: Below a critical value of the
coupling constant $g$, the trace distance as a function of $\beta$
exhibits an initial peak and then goes down to zero, see also
Fig.~(\ref{fig:3}.a); above this critical $g$ it keeps growing to an
asymptotic value different from zero, which we will discuss later on,
as can be seen from Fig.~(\ref{fig:3}.b). On the other hand, the
dependence of the trace distance on $g$ for a fixed value of $\beta$
shows some oscillations after a sudden growth which occurs at the
critical $g$, see Figs.~(\ref{fig:2}) and (\ref{fig:3}.d). Quite
remarkably, this means that the total amount of correlations of the
Gibbs state can decrease with increasing coupling constant,
as clearly observed in Fig.~(\ref{fig:3}.d).

\begin{figure}[h]
\begin{center}
\includegraphics[scale=1.]{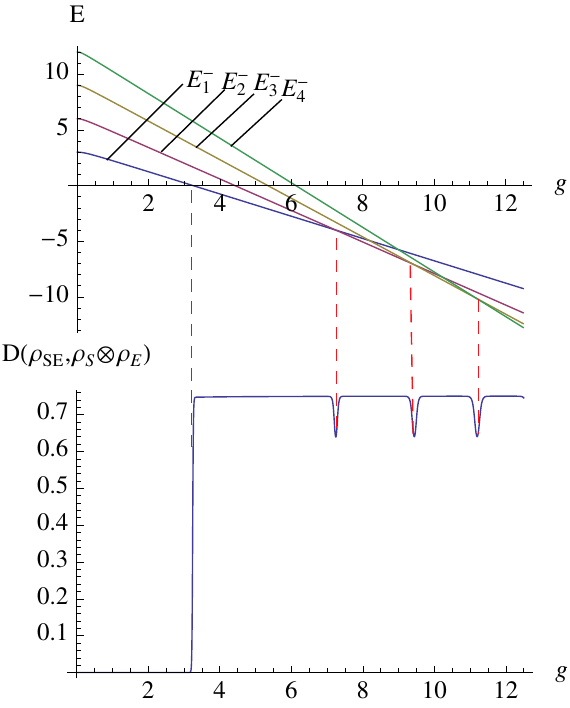}
\end{center}
\caption{\label{fig:4} (Color online) (Top) Plot of the first
energy eigenvalues $E^-_1, E^-_2, E^-_3, E^-_4$ given by
Eq.~(\ref{eq:eigenvalues}) as functions of $g$, $E^-_0$ coincides
with the $x$-axis. (Bottom) Plot of the correlations of the Gibbs
state as a function of $g$ for $\beta = 100$, i.e., for
approximately zero temperature; the other values are the same as
in Fig.~(\ref{fig:2}). The critical values of the
correlations as a function of $g$ exactly correspond to the level
crossing points: When $E^-_0=E^-_1$ there is a sudden increase
and at the subsequent points the dips occur.  For this value of
$\beta$ the behavior described by the exact expression is well
approximated by Eq.~(\ref{eq:correlip1}) between the dips and by
Eq.~(\ref{eq:peaks}) at the dips.}
\end{figure}

The above features can be explained considering that the trace
distance $D(\rho_{SE},\rho_S\otimes\rho_E)$ quantifies the
correlations of the Gibbs state $\rho_{SE}$ and that the limit
$\beta \rightarrow \infty$ corresponds to the limit of zero
temperature, where the Gibbs state reduces to the ground state of
the Hamiltonian $H$. If all the eigenvalues given by
Eq.~(\ref{eq:eigenvalues}) are non-negative the ground state is
$|\Phi^-_0\rangle=|0,0\rangle$ with eigenvalue zero. Of course,
this is a product state and, therefore, the correlations of the
Gibbs state approach zero for $\beta\to\infty$. This is what
happens below the critical $g$. However, according to
the level crossing described in Fig.~(\ref{fig:4}), the
Hamiltonian given by Eq.~(\ref{eq:h}) has negative eigenvalues for
larger values of the coupling constant $g$. In fact, it is easy to
see from Eq.~(\ref{eq:eigenvalues}) that if
\begin{equation}
  g > \bar{g}_1 \equiv \sqrt{\omega^2 + \omega \Delta}  \label{eq:criticalg}
\end{equation}
then $E_1^- < 0$ and, therefore, $|0, 0 \rangle$ is no longer the
ground state. Thus, we can then identify $\bar{g}_1$ as the
previously mentioned critical value of $g$, since for larger
values the lowest energy state is $| \Phi^-_1 \rangle$ which is an
entangled state according to Eq.~(\ref{eq:dressed}) with
correlations $a_1^2 b_1^2 + a_1 b_1$ different from zero. But
looking at the dependence of the different eigenvalues $E_n^-$ on
the coupling constant $g$, see Fig.~(\ref{fig:4}), we can see that
there is another critical point, let us call it $\bar{g}_2$, where
$E^-_2 ( \bar{g}_2) = E^-_1 ( \bar{g}_2)$ and after which $E^-_2
(g) < E^-_1 (g)$, i.e., $| \Phi_2^- \rangle$ becomes the lowest
energy state. We then have another value $\bar{g}_3$ for which
$E^-_3 ( \bar{g}_3) = E^-_2 ( \bar{g}_3)$, so that for stronger
couplings $| \Phi^-_3 \rangle$ becomes the new ground state, and
so on. Between two successive critical values $\bar{g}_i$ and
$\bar{g}_{i + 1}$ the ground state of the Hamiltonian is $|
\Phi_i^- \rangle$, whose correlations according to
Eq.~(\ref{eq:limrrpure}) are given by
\begin{equation}
  D\left(\rho_{SE},\rho_S\otimes\rho_E\right) =
  a_i^2 b_i^2 + a_i b_i = \frac{g^2}{\Delta^2 + 4 g^2}
  + \sqrt{\frac{g^2}{\Delta^2 + 4 g^2}} .  \label{eq:correlip1}
\end{equation}
This expression characterizes the asymptotic value of the
correlations in the Gibbs state for $\beta\to\infty$ and for $g$
between $\bar{g}_i$ and $\bar{g}_{i + 1}$. We note that
$D(\rho_{SE},\rho_S\otimes\rho_E)$ approaches the value
$\frac{3}{4}$ if we also let $g\to\infty$. As is shown in the
Appendix, this asymptotic value corresponds in fact to the maximal
possible value of the correlations for the present model.

We see from Fig.~(\ref{fig:4}) that for small temperatures the
correlations in the Gibbs state exhibit a dip at every $\bar{g}_i$
with $i>1$. Again, this feature can be explained considering the
ground level of the Hamiltonian given by Eq.~(\ref{eq:h}). For $g
= \bar{g}_i$ the eigenspace of the lowest energy level is two-fold
degenerate since $E_i^- ( \bar{g}_i) = E_{i - 1}^- ( \bar{g}_i)$
and the Gibbs state reduces to
\begin{eqnarray}
 && \frac{1}{2} \left( | \Phi_{i - 1}^- \rangle \langle \Phi^-_{i - 1} | + |
  \Phi_i^- \rangle \langle \Phi^-_i | \right)  \label{eq:grdeg}
\end{eqnarray}
where, again we have ordered the elements of the basis as $\left\{
|1, i - 2 \rangle, |0, i - 1 \rangle, |1, i - 1 \rangle, |0, i
\rangle \right\}$. Equation~(\ref{eq:grdeg}) can be directly
obtained from Eq.~(\ref{eq:gibbsmat}), observing that for $\beta
\rightarrow \infty$ the only non-negligible terms are those
involving the exponentials of $\beta E^-_{i - 1}$ or $\beta
E_i^-$. Calculating now the corresponding product state and
proceeding as done to obtain Eq.~(\ref{eq:tdgibbs}), or directly
taking the limit of this equation for $\beta\rightarrow\infty$ and
$g=\bar{g}_i$, one finds an explicit expression for the
correlations of the mixed state given by Eq.~(\ref{eq:grdeg}):
\begin{eqnarray}
 && D \left( \rho^{}_{SE}, \rho^{}_S \otimes \rho^{}_E \right)\nonumber\\
 &&=  \frac{1}{2} \left[ \alpha
 + \frac{1}{2} | \gamma_1 + \delta_1 + \sqrt{(\gamma_1 -
  \delta_1)^2 + 4 \varepsilon^2_1} |\right.\nonumber\\
  &&\left.+ \frac{1}{2} | \gamma_1 + \delta_1 -
  \sqrt{(\gamma_1 - \delta_1)^2 + 4 \varepsilon^2_1} | \right. \nonumber\\
  &  &  + \frac{1}{2} | \gamma_2 + \delta_2 + \sqrt{(\gamma_2 -
  \delta_2)^2 + 4 \varepsilon^2_2} |\nonumber\\
  && \left.+ \frac{1}{2} | \gamma_2 + \delta_2 -
  \sqrt{(\gamma_2 - \delta_2)^2 + 4 \varepsilon^2_2} | + \chi \right] ,
  \label{eq:peaks}
\end{eqnarray}
where
\begin{eqnarray}
 && \alpha =  \frac{b_{i - 1}^2}{4} \left( a_{i - 1}^2 + a_i^2 \right);\,\,\,\,
  \gamma_1  =  \frac{b_{i - 1}^2}{2} - \frac{b_{i - 1}^2}{4} \left( b_{i -
  1}^2 + b_i^2 \right); \nonumber\\
 && \delta_1  =  \frac{a_{i - 1}^2}{2} - \frac{1}{4} \left( a_{i - 1}^2 +
  a_i^2 \right) \left( a_{i - 1}^2 + b_i^2 \right);\,\,\,
  \varepsilon_1  =  - \frac{a_{i - 1} b_{i - 1}}{2}; \nonumber\\
  &&\gamma_2  =  \frac{b_i^2}{2} - \frac{1}{4} \left( b_{i - 1}^2 + b_i^2
  \right) \left( a_{i - 1}^2 + b_i^2 \right);\,\,\,\,\varepsilon_2  =  - \frac{a_i b_i}{2}; \nonumber\\
 && \delta_2 =  \frac{a_i^2}{2} - \frac{a_i^2}{4} \left( a_{i - 1}^2 + a_i^2
  \right); \,\,\,\, \chi  =  \frac{a_i^2}{4} \left( b_{i - 1}^2 + b_i^2 \right).
  \label{eq:coeffi}
\end{eqnarray}
From the explicit evaluation of Eq.~(\ref{eq:correlip1}) and
Eq.~(\ref{eq:peaks}) for the different values of $i$, one can see
that indeed the total amount of correlations of the mixed state
given by Eq.~(\ref{eq:grdeg}) is smaller than the correlations of
the dressed states $| \Phi_{i - 1}^- \rangle$ and $| \Phi^-_i
\rangle$ giving its decomposition, which explains the emergence of
the dips. Note however that the correlation measure
given by $D(\rho_{SE},\rho_S\otimes \rho_E)$ is not a convex
function on the space of physical states.

The above arguments are summarized in Fig.~(\ref{fig:4}). They
explain the behavior of the correlations in the Gibbs state for
small temperatures, i.e., for $\beta \rightarrow \infty$. The
effect of finite temperatures is to smoothen the dependence on
$g$, as can be seen in Fig.~(\ref{fig:4}), (\ref{fig:3}.d) and
(\ref{fig:3}.c), such that the sudden increase at $g = \bar{g}_1$
is less sharp and that the subsequent dips turn into oscillations
which are more and more suppressed as the temperature increases.
This behavior is due to the fact that at finite temperature the
Gibbs state has a non-vanishing admixture of
$|\Phi^{-}_{1}\rangle$ for values of $g$ which are smaller than
$\bar{g}_1$ and, hence, the increase of the correlations starts
before $g = \bar{g}_1$ and is less sharp, as can be seen from
Figs.~(\ref{fig:4}) and (\ref{fig:3}.d). Moreover, as a
consequence of finite temperatures, the Gibbs state is a mixed
state even between the critical values $\bar{g}_i$, such that its
correlations become smaller than in the zero temperature limit,
which leads to a suppression of the oscillations.

\subsection{Time evolution of the trace distance}

The analysis performed so far concerns the correlations of the
initial Gibbs state, i.e., the upper bound of the trace distance
between the reduced state $\rho^1_S (t)$, evolving from an initial
total Gibbs state, and the reduced state $\rho^2_S (t)$, evolving
from the corresponding product state, according to
Eq.~(\ref{eq:llaine}). We will now investigate the dynamics of the
trace distance $D(\rho^1_S(t),\rho^2_S(t))$ and analyze, in
particular, the dependence of the supremum of this function on the
coupling constant and the temperature. As discussed before (see
Sec.~\ref{Sec:D_Dyn}), the behavior of the trace distance between
$\rho^1_S(t)$ and $\rho^2_S(t)$ expresses the effect of the
initial correlations in the resulting dynamics. Moreover, its
supremum as a function of time quantifies the amount of
information which could not be initially retrieved by measurements
on the reduced system only, but becomes accessible in the
subsequent dynamics, thus making the two reduced states
$\rho^1_S(t)$ and $\rho^2_S(t)$ more distinguishable.

\begin{figure}[h]
\begin{center}
\includegraphics[scale=0.7]{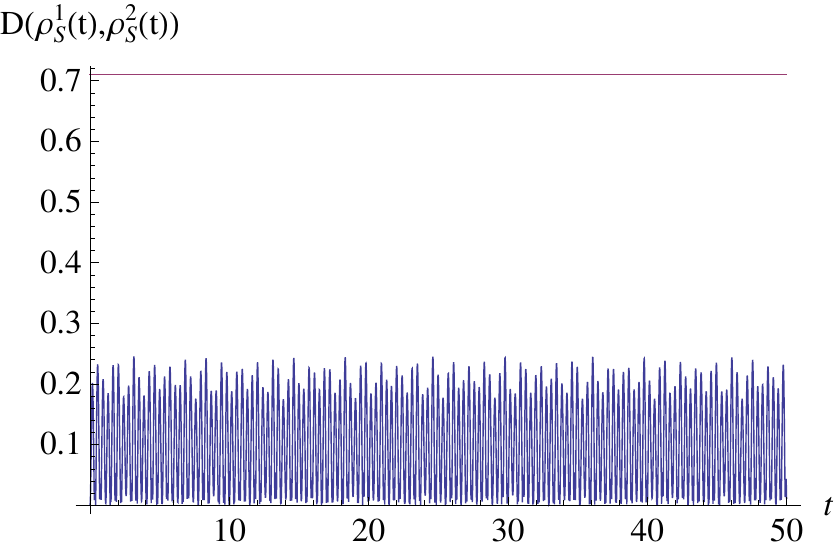}
\caption{\label{fig:5} (Color online) The trace distance
$D(\rho^1_S(t),\rho^2_S(t))$ as a function of time according to
Eq.~(\ref{eq:maxv}); $\rho^i_S (t)$ is the state of the reduced
system at time $t$ obtained from an initial total state
$\rho^i_{SE}$, where $\rho^1_{SE}$ is the Gibbs state given by
Eq.~(\ref{eq:gibbs}) and $\rho^2_{SE}$ is the corresponding
product state. The upper horizontal line represents
the bound given by the right-hand side of the inequality
(\ref{eq:llaine}) which has been determined by
Eq.~(\ref{eq:tdgibbs}). Parameters: $\omega=3$, $\Delta=0.5$,
$g=6$ and $\beta=5$.}
\end{center}
\end{figure}

Taking as initial state $\rho^1_{SE}$ the Gibbs state
given by Eq.~(\ref{eq:gibbs}) and $\rho^2_{SE}$ as the
corresponding product of its marginals, we have
$\rho^1_S(t)=\rho^1_S(0)$ since the Gibb state is invariant under
the time evolution, and $\rho_S^1(0)=\rho^2_S (0)$ because the
corresponding open system initial states are identical. Thus,
exploiting Eq.~(\ref{eq:red}) we obtain the following explicit
expression for the trace distance,
\begin{eqnarray}
  D(\rho^1_S (t),\rho^2_S (t))& = &\left| \left( \rho_{00} - 1 \right) \sum_n
  (n + 1) \rho^{nn} |d_{n + 1} (t) |^2 \right.\nonumber\\ &&\left.+ \rho_{00} \sum_n n \rho^{nn}
  |d_n (t) |^2 \right|.  \label{eq:maxv}
\end{eqnarray}
For fixed values of the parameters characterizing the dynamics
this expression describes a superposition of periodic functions
with incommensurable periods, i.e., an almost periodic function as
already encountered in Sec.~\ref{sec:ex}. An example for the trace
distance dynamics is shown in Fig.~(\ref{fig:5}). The trace
distance starts growing already at the initial time and further
oscillates with time, according to the almost periodic behavior
described by Eq.~\eqref{eq:maxv}.

As mentioned already the time dependence of the trace
distance is solely due to the time evolution of the product state
constructed from the marginals of the Gibbs state since the latter
is invariant under the dynamics. It is the comparison between the
two different reduced system states, namely between the states
$\rho^1_S (t)=\rho^1_S(0)$ and $\rho^2_S(t)$, which allows to
obtain information initially not accessible with measurement on
the reduced system only, and which enables the detection of
correlations in the initial Gibbs state.

The supremum of the trace distance in Fig.~(\ref{fig:5}) is
substantially smaller than the corresponding bound of
Eq.~(\ref{eq:llaine}).  For large values of $\beta$ and $g$ the
supremum can be estimated as follows. If the temperature goes to zero
the Gibbs state approaches the projection onto the ground state which
is given by $| \Phi^-_k \rangle\langle \Phi^-_k |$ for a fixed $k$,
depending on the value of the coupling constant $g$.  We suppose that
$g$ is different from the critical values $\bar{g}_i$. This implies
$\rho_{00}=a_k^2$, $\rho_{11}=b_k^2$, together with
$\rho^{mm}=\delta_{m,k}a_k^2+\delta_{m,k-1}b_k^2$.  For large values
of $g$, which implies large values of $k$, we have $a_k \approx b_k
\approx {1}/{\sqrt{2}}$. Employing further Eq.~(\ref{eq:cd}), one thus
obtains the estimate
\begin{equation} \label{eq:1000}
 D(\rho^1_S (t),\rho^2_S (t)) \approx \frac{1}{4} \left|
 \sin(2\sqrt{k}g t) \sin (gt/\sqrt{k}) \right|.
\end{equation}
This shows that for large $\beta$ and $g$ the trace distance is
bounded from above by $\frac{1}{4}$.

\begin{figure}[h]
\includegraphics[scale=0.71]{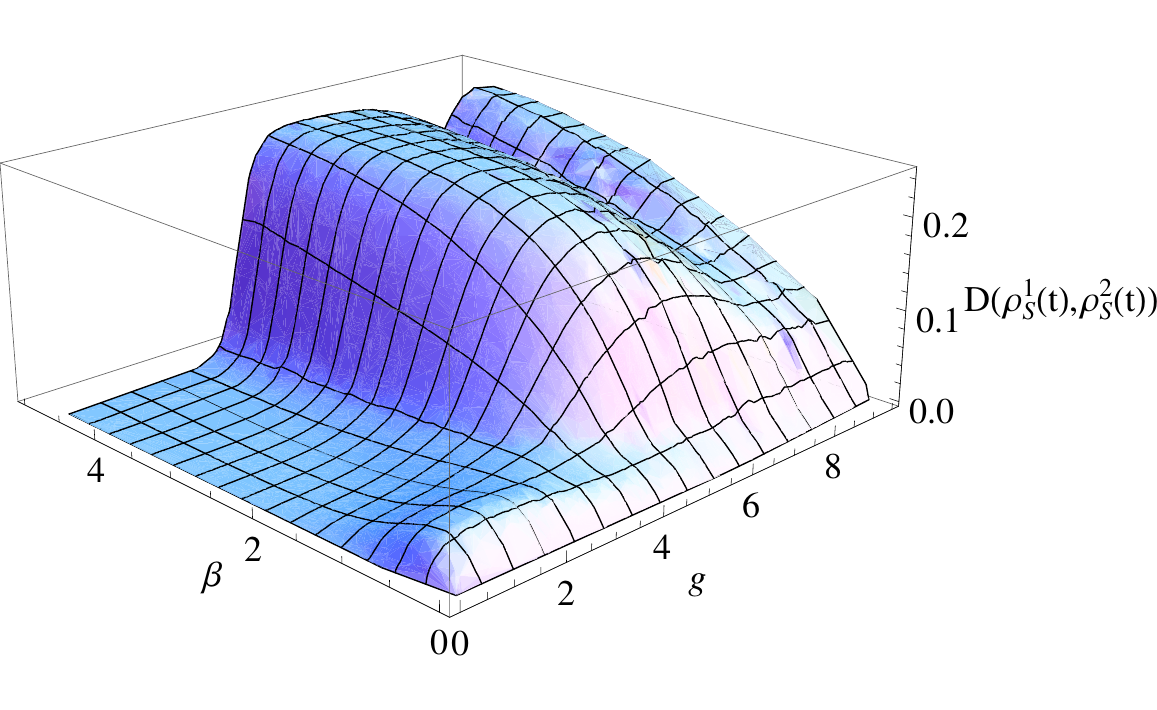}
\caption{\label{fig:6} (Color online) The supremum of
$D(\rho^1_S(t),\rho^2_S(t))$ as a function of time versus the
coupling constant $g$ and the inverse temperature $\beta$;
$\rho^1_S (t)$ is obtained from an initial total Gibbs state,
$\rho_S^2 (t)$ from the corresponding product state,
$D(\rho^1_S(t),\rho^2_S(t))$ is calculated according to
Eq.~(\ref{eq:maxv}). Parameters: $\omega=3$, $\Delta=0.5$.}
\end{figure}

\begin{figure}[h]
\begin{center}
\includegraphics[scale=0.44]{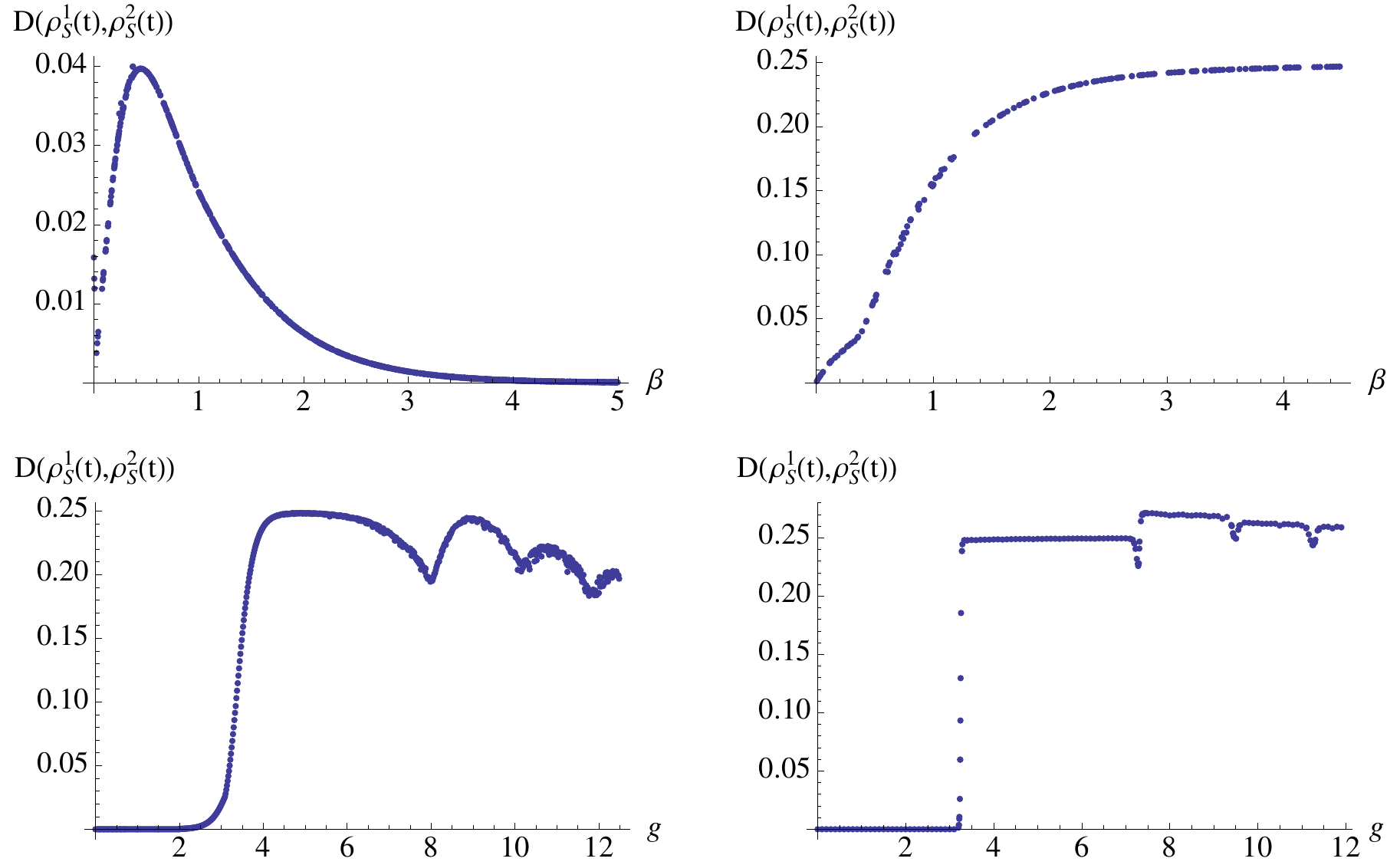}
\end{center}
\caption{\label{fig:7} (Color online) (a, b, c, d From top left to
bottom right) The same as Fig.~(\ref{fig:6}) but for parameters
$g=1.7$, $g = 5.5$, $\beta = 5$ and $\beta = 100$, respectively.
For $\beta=100$, i.e., approximately zero temperature, the dips
occur at the same values as the corresponding dips of the bound,
see Fig.~(\ref{fig:4}). For the case of finite temperature the dips
are not suppressed, but they are shifted towards larger values of $g$.}
\end{figure}

Figure (\ref{fig:6}) shows how the supremum of $D
(\rho^1_S(t),\rho^2_S(t))$ behaves as a function of the coupling
constant $g$ and the inverse temperature $\beta$, keeping fixed
$\omega$ and $\Delta$.  Exactly as for the correlations of the
Gibbs state [compare with Fig.~(\ref{fig:2})], we observe two
qualitatively different kinds of behavior as a function of
$\beta$, for a fixed value of $g$. Below a critical $g$ the
supremum of the trace distance passes through maximum and then
tends to zero; above the critical value it tends monotonically to
an asymptotic value which is close to the estimate of
$\frac{1}{4}$ determined above, as illustrated in
Figs.~(\ref{fig:7}.a) and (\ref{fig:7}.b). Moreover, considering
the supremum of the trace distance as a function of $g$ for fixed
$\beta$, after a sudden growth at the first critical $g$ it
exhibits some oscillations analogous to those of the bound.
Comparing Figs.~(\ref{fig:4}) and (\ref{fig:7}), we see that in
the limit of zero temperature the bound and the true supremum of
the trace distance both show a sudden increase and subsequent dips
at the same values of the coupling constant $g$.  This behavior
can be explained by recalling the dependence of the energy
spectrum of the Hamiltonian as a function of $g$ in
Fig.~(\ref{fig:4}).  At zero temperature the Gibbs state reduces
to the ground level of the Hamiltonian. The discontinuous change
in the ground level with varying $g$, i.e. the transition from
$|\Phi_i\rangle$ to
$(|\Phi_i\rangle+|\Phi_{i+1}\rangle)/\sqrt{2}$, implies a
discontinuous change in the bound as well as in the supremum of
the trace distance, thus leading to the dips appearing in
Fig.~(\ref{fig:4}) and Fig.~(\ref{fig:7}).  In fact, apart from
fixing the bound at the r.h.s. of Eq.~(\ref{eq:llaine}), the Gibbs
state determines both reduced states $\rho^1_S(t)$ and
$\rho^2_S(t)$, arising from the initial total states $\rho_{SE}$
and $\rho_S \otimes \rho_E$ respectively.  Relying on
Eq.~(\ref{eq:eqvt}) one can see that for $\Delta=0$ the supremum
of the trace distance is simply given by {1}/{4} for an initial
correlated state $\rho_{SE}=|\Phi_i\rangle\langle\Phi_i|$, for any
$i>0$.  This means that for zero detuning the supremum of the
trace distance dynamics as a function of $g$ at zero temperature
takes the constant value $1/4$, except at $g=\bar{g}_i$ where the
dips occur.  The effect of a finite temperatures is slightly
different for the bound and the supremum of the trace distance
dynamics: With growing temperature the dips of the bound turn into
oscillations which are more and more suppressed, but they occur at
the same values of $g$. On the contrary, the dips of the true
supremum, and its sudden increase as well, are not suppressed, but
do change position,
occurring at larger values of $g$.

\section{Conclusions}
We have studied the influence of initial correlations between
system and reservoir on the dynamics of an open quantum system by
means of the trace distance, considering the paradigmatic and
exactly solvable model provided by the Jaynes-Cummings
Hamiltonian. First, we have analyzed the amount of correlations in
the Gibbs state $\rho_{SE}$ as it is quantified by $D(\rho_{SE},
\rho_S\otimes\rho_E)$, where $\rho_S\otimes\rho_E$ denotes the
product state arising from the marginals of $\rho_{SE}$. The exact
analytical expression of the latter quantity describes a
non-monotonic behavior as a function of both the coupling constant
and the temperature characterizing the dynamics. The
same behavior is found for the supremum of the trace distance
between the open system states $\rho^1_S(t)$ and $\rho^2_S(t)$
which evolve from $\rho_{SE}$ and $\rho_S\otimes\rho_E$,
respectively. This enabled us to establish a clear connection
between the correlation properties of the Gibbs state and basic
features of the subsequent open system dynamics, in particular,
the amount of information which is initially inaccessible for the
open system and which is uncovered during its time evolution. The
dynamical behavior of the trace distance between the reduced
system states $\rho^1_S(t)$ and $\rho^2_S(t)$ thus provides a
witness for the correlations of the total system's initial state.

As we have shown for the case at hand, at zero temperature sudden
changes in the supremum over time of the trace distance can be
traced back uniquely to discontinuous changes in the structure of
the total system's ground state and to its degree of entanglement
which, in turn, is caused by crossings of the energy levels of the
total system Hamiltonian. It is important to remark that, as we
have demonstrated, clear signatures of these discontinuities are
still present at finite temperatures. Note that to reconstruct the
trace distance dynamics, in order to detect correlation properties
of the ground state, one only needs to follow the evolution of the
open system state obtained from the initial total product state
$\rho_S\otimes\rho_E$.

The bound given by the right-hand side of Eq.~(\ref{eq:llaine}) is
able to represent qualitatively the non-trivial behavior of the
maximum of the trace distance between $\rho^1_S (t)$ and $\rho^2_S
(t)$, as a function of the different parameters characterizing the
Hamiltonian and the temperature. While for the sudden transition
between the two different asymptotic regimes as a function of
$\beta$ it is clear that the effective maximum of $D (\rho^1_S
(t), \rho^2_S (t))$ has to reproduce the behavior of the bound, it
is quite remarkable that also in the second case, where the bound
is sensibly different from the effective maximum and from zero,
both these quantities show an analogous non-monotonic behavior.
Finally, we note that it must be expected that the general features
found here for the correlated Gibbs state hold true also for other
correlated initial states, e.g., for correlated non-equilibrium
stationary states, as long as the latter involve discontinuous,
qualitative changes under the variation of some system
parameters.

\begin{acknowledgments}
Financial support by the Ministero dell'Istruzione,
dell'Universit\`a e della Ricerca (MIUR), under PRIN2008, by the
Academy of Finland (project 133682), and by the Magnus Ehrnrooth
Foundation are gratefully acknowledged.
\end{acknowledgments}

\appendix*

\section{General bound for the correlations of a quantum state}

Throughout the article we have used the quantity
$D(\rho_{SE},\rho_{S}\otimes\rho_E)$ as a measure for the total amount
of correlations of given state $\rho_{SE}$. On the ground of extensive numerical simulations we conjecture
that this quantity
satisfies the inequality
\begin{equation}
 D(\rho_{SE},\rho_{S}\otimes\rho_E) \leq 1 - \frac{1}{N^2},
 \label{eq:strsub2}
\end{equation}
where $N$ denotes the minimum of the dimensions of $\mathcal{H}_S$
and $\mathcal{H}_E$. For the example studied in this paper we have
$N=2$ and, hence, $D(\rho_{SE},\rho_S\otimes\rho_E)\leq
\frac{3}{4}$.

To our knowledge there exists no general mathematical proof for
the inequality \eqref{eq:strsub2}. However, one can easily prove
that this inequality is saturated if
$\rho_{SE}=|\psi\rangle\langle\psi|$ is a pure, maximally
entangled state. To show this we first note that for a maximally
entangled state vector $|\psi\rangle$ the marginal states are
given by $\rho_S=P_S/N$ and $\rho_E=P_E/N$, where $P_S$ and $P_E$
are the projections onto the subspaces of $\mathcal{H}_S$ and
$\mathcal{H}_E$, respectively, which are spanned by the local
Schmidt basis vectors with nonzero Schmidt coefficients. Hence,
$D(\rho_{SE},\rho_S\otimes\rho_E)$ is given by $\frac{1}{2}$ times
the sum of the absolute eigenvalues of the operator
\begin{equation}
 X = |\psi\rangle\langle\psi|-\frac{1}{N^2}P_S\otimes P_E.
 \label{eq:chi}
\end{equation}
Obviously, $|\psi\rangle$ is an eigenvector of $X$ corresponding
to the eigenvalue $1-1/N^2$. Moreover, all vectors which are
perpendicular to $|\psi\rangle$ and belong to the support of
$P_S\otimes P_E$ are eigenvectors of $X$ with the eigenvalue
$-1/N^2$. Thus, $X$ has one non-degenerate eigenvalue $1-1/N^2$,
and one eigenvalue $-1/N^2$ which is $(N^2-1)$-fold degenerate,
while all other eigenvalues of $X$ are zero. Therefore we have
\begin{equation}
 D(\rho_{SE},\rho_S\otimes\rho_E) =
 \frac{1}{2}\left[1-\frac{1}{N^2}+(N^2-1)\frac{1}{N^2}\right] =
1-\frac{1}{N^2},
\end{equation}
which proves the claim.


\begin{thebibliography}{10}

  \bibitem{Breuer2007}H.-P.~Breuer and F.~Petruccione,
  \tmtextit{The Theory of Open Quantum Systems}  (Oxford
  University Press, Oxford, 2007).

  \bibitem{Gorini1976a}V.~Gorini, A,~Kossakowski, and E.~C.~G.
  Sudarshan, J.~Math. Phys.  \textbf{17}, 821 (1976).

  \bibitem{Lindblad1976a}G.~Lindblad, Comm. Math. Phys.
   \textbf{48}, 119 (1976).

  \bibitem{Daffer2004a}S.~Daffer, K.~ W\'odkiewicz, J.~D.~
  Cresser, and J.~K.~McIver, Phys. Rev.~A  \textbf{70}, 010304 (2004).

  \bibitem{Budini2004a}A.~A.~Budini,
  Phys. Rev.~A  \textbf{69}, 042107 (2004).

  \bibitem{Shabani2005a}A.~Shabani and D.~A. Lidar,
  Phys. Rev.~A  \textbf{71}, 020101 (2005).

  \bibitem{Maniscalco2006a}S.~Maniscalco and F.~Petruccione,
  Phys.
  Rev.~A  \textbf{73}, 012111 (2006).

  \bibitem{Vacchini2008a}B.~Vacchini,  Phys. Rev.~A
   \textbf{78}, 022112 (2008).

  \bibitem{Ferraro2008a}E.~Ferraro, H.-P. Breuer, A.~Napoli, M.~A.
  Jivulescu, and A.~Messina, Phys.
  Rev.~B  \textbf{78}, 064309 (2008).

  \bibitem{Kossakowski2008a}A.~Kossakowski and R.~Rebolledo,
   Open Syst. Inf. Dyn.  \textbf{15}, 135 (2008).

  \bibitem{Kossakowski2009a}A.~Kossakowski and R.~Rebolledo,
  Open Syst. Inf. Dyn.  \textbf{16}, 259 (2009).

  \bibitem{Breuer2008a}H.-P.~ Breuer and B.~ Vacchini,
 Phys. Rev.
  Lett.  \textbf{101}, 140402 (2008).

  \bibitem{Breuer2009a}H.-P.~Breuer and B.~Vacchini,
   Phys. Rev.~E  \textbf{79}, 041147 (2009).

  \bibitem{Piilo2008a}J ~Piilo, S.~Maniscalco, K.~H\"ark\"onen, and
  K.-A.~Suominen,
  Phys. Rev. Lett.  \textbf{100}, 180402 (2008).

  \bibitem{Piilo2009a}J.~Piilo, K.~H\"ark\"onen, S.~Maniscalco, and K.-A.
  Suominen, Phys. Rev.~A  \textbf{79}, 062112 (2009).

   \bibitem{Breuer2009d}H.-P.~Breuer and J.~Piilo,
  Europhys. Lett.
   \textbf{85}, 50004 (2009).

  \bibitem{Breuer2009c}H.-P.~Breuer, E.-M.~Laine, and J.~Piilo,
  Phys. Rev. Lett.
   \textbf{103}, 210401 (2009).

  \bibitem{Laine2010a}E.-M.~Laine, J.~Piilo, and H.-P.~Breuer,
  Phys. Rev.~A  \textbf{81}, 062115 (2010).

  \bibitem{Vacchini2010b}B.~Vacchini and H.-P.~Breuer,
  Phys. Rev.~A  \textbf{81}, 042103 (2010).

  \bibitem{Smirne2010a}A.~Smirne and B.~Vacchini, Phys. Rev.~A  in press (2010).

  \bibitem{Chruscinski2010a}D.~Chru\'{s}ci\'{n}ski and A.~Kossakowski,
  Phys. Rev. Lett.  \textbf{104}, 070406 (2010).

  \bibitem{Chruscinski2010b}D.~Chru\'{s}ci\'{n}ski, A.~Kossakowski, and
  S.~Pascazio,  Phys. Rev.~A  \textbf{81}, 032101 (2010).

  \bibitem{Mazzola2010a}L.~Mazzola, E.-M.~Laine, H.-P.~Breuer, S.~Maniscalco, and J.~Piilo, Phys.
  Rev.~A  \textbf{81}, 062120 (2010).

  \bibitem{Pechukas1994a}P.~Pechukas, Phys. Rev. Lett.  \textbf{73}, 1060
  (1994).

  \bibitem{Alicki1995a}R.~Alicki, Phys. Rev. Lett.
   \textbf{75}, 3020 (1995).

  \bibitem{Uchiyama2010a}C.~Uchiyama and M.~Aihara, Phys.~Rev.~A.  \textbf{82}, 044104 (2010).

  \bibitem{Smirne2010b}A.~Smirne and B.~Vacchini,
   Phys.
  Rev.~A  \textbf{82}, 022110 (2010).

  \bibitem{vanWonderen2000a}A.~J.~van~Wonderen and K.~Lendi,
   J.~Phys.~A: Math. Gen.  \textbf{33}, 5757 (2000).

  \bibitem{Stelmachovic2001a}P.~\v{S}telmachovi\v{c} and V.~Bu\v{z}ek,
   Phys.
  Rev.~A  \textbf{64}, 062106 (2001).

  \bibitem{Jordan2004a}T.~F.~Jordan, A.~Shaji, and E.~C.~G.~Sudarshan,
  Phys. Rev.~A  \textbf{70}, 052110 (2004).

  \bibitem{Rodriguez2008a}C.~A.~Rodr\'{\i}guez-Rosario, K.~Modi, A.~Kuah,
  A.~Shaji, and E.~C.~G. Sudarshan, J.~Phys.~A: Math. Gen.
   \textbf{41}, 205301 (2008).

  \bibitem{Carteret2008a}H.~A.~Carteret, D.~R.~Terno, and K.~\.Zyczkowski,
   Phys. Rev.~A  \textbf{77}, 042113 (2008).

  \bibitem{Shabani2009a}A.~Shabani and D.~A. Lidar, Phys. Rev. Lett.  \textbf{102}, 100402 (2009).

  \bibitem{Laine2010b}E.-M.~Laine, J.~Piilo, and H.-P.~Breuer (2010), arXiv:1004.2184 [quant-ph].

  \bibitem{Dajka2010a}J.~Dajka and J.~\L uczka,  Phys. Rev.~A  \textbf{82}, 012341 (2010).

  \bibitem{Ruskai1994a}M.~B.~Ruskai, Rev. Math. Phys.  \textbf{6}, 1147 (1994).

  \bibitem{Gilchrist2005a}A.~Gilchrist, N.~K.~Langford, and M.~A.~Nielsen,
  Phys. Rev.~A  \textbf{71}, 062310 (2005).

  \bibitem{Puri2001}R.~R.~Puri, \tmtextit{Mathematical Methods
  of Quantum Optics}  (Springer, Berlin, 2001).

  \bibitem{Corduneanu1989}C.~Corduneanu. \tmtextit{Almost
  Periodic Functions} (Chelsea Publishing Company, New York, 1989).

   \bibitem{Note1}

Due to the incommensurability of the frequencies,
there is no time $t$ at which $D(\rho ^1_S(t),\rho ^2_S(t))$
attains the supremum.

  \bibitem{Meystre1991a}P.~Meystre and M.~Sargent III.
  \tmtextit{Elements of Quantum Optics}
  (Springer-Verlag, Berlin, 1991).

\end{thebibliography}
\end{document}